\documentclass[12pt]{article}

\usepackage{feynmp}
\usepackage{tikz-feynman}
\tikzfeynmanset{compat=1.1.0}
\usepackage{slashed}
\usepackage{subcaption}
\usepackage{amssymb,amsmath}

\begin{document}

\begin{titlepage}

\begin{flushright}
arXiv:2110.08646
\end{flushright}
\vskip 2.5cm

\begin{center}
{\Large \bf Renormalization Scheme Dependence of $\beta$-Functions\\
In Lorentz-Violating Quantum Field Theory}
\end{center}

\vspace{1ex}

\begin{center}
{\large Sapan Karki and Brett Altschul\footnote{{\tt altschul@mailbox.sc.edu}}}

\vspace{5mm}
{\sl Department of Physics and Astronomy} \\
{\sl University of South Carolina} \\
{\sl Columbia, SC 29208} \\
\end{center}

\vspace{2.5ex}

\medskip

\centerline {\bf Abstract}

\bigskip

Effective quantum field theories that allow for the possibility of Lorentz symmetry violation
can sometimes also include redundancies of description in their Lagrangians.
Explicit calculations in a Lorentz-violating generalization
of Yukawa theory show that when this kind of redundancy exists, different renormalization schemes
may lead to different expressions for the renormalization group
$\beta$-functions, even at only one-loop order. However, the renormalization group
scaling of physically observable quantities appears not to share this kind
of scheme dependence.

\bigskip

\end{titlepage}

\newpage

\section{Introduction}

The special theory of relativity is one of the cornerstones of modern physics, but
ever since Einstein introduced it in 1905, there has always
been interest in asking whether the Lorentz symmetry underlying special relativity is truly exact,
or whether it is just a highly useful approximation.
Apparent symmetries that eventually turn out to be merely approximate have played extremely important
roles in the development of the standard model of particle physics, and it is natural to wonder whether
Lorentz symmetry may represent a similar case.
Over the last quarter century, owing to developments in effective field theory (EFT) techniques, it has become
possible to describe possible violations of rotation invariance and Lorentz boost invariance in a
systematic way. Along with this theoretical development, there came a burst of experimental interest
in Lorentz symmetry tests---because it became apparent that there were a great many possible forms of
Lorentz violation that had hardly been previously constrained at all. The
renewed experimental searches have, thus far,
not identified any particularly compelling evidence in favor of Lorentz violation. However, it remains
a significant area of experimental research, because we know that if Lorentz violation is ever really
demonstrated to exist, that will be a colossally important discovery, opening up whole new avenues for
the study of physics at the most fundamental levels.

The most general local effective field theory for describing Lorentz-violating modifications to 
the physics of the well-known fields that are part of the standard model of particle physics has already been
laid out in detail. This theory, which is called the standard model extension (SME), is also
capable of describing stable, unitary, local forms of CPT violation involving
standard model fields, since there can be connections between CPT violation and
Lorentz violation~\cite{ref-greenberg}---although the two occur separately in certain
theories~\cite{ref-chaichian1,ref-dutsch,ref-chaichian2}.
The SME is a quantum field theory (QFT), whose action
contains operators build out of the fermion and boson fields of the standard
model~\cite{ref-kost1,ref-kost2}. The minimal SME is the sector of the SME that is
expected to be renormalizable; it contains only the finite number of local, Hermitian, gauge-invariant
operators that are of
mass dimension four or less.
The terms in the minimal SME Lagrangian are actually similar in structure to those found in the usual standard
model---the key difference being that the SME operators may have uncontracted Lorentz indices.
The minimal SME is very often the most
useful test theory framework for evaluating the results of experimental Lorentz tests.

Really understanding a QFT entails understanding the role that quantum corrections play, and this,
in turn, means understanding renormalization.
Quite a bit of progress has been made toward describing the renormalization of the minimal SME,
particularly at one loop order~\cite{ref-kost3,ref-berr,ref-collad-3,
ref-collad-2,ref-collad-1,ref-gomes,
ref-anber,ref-ferrero3,ref-brito1}. And yet,
some questions related to the renormalization of the minimal SME are still outstanding, and this is
particularly true in relation to some of the less commonly discussed sectors of the theory. For example, the
explicit calculations needed for the one-loop renormalization of a Lorentz-violating
gauge theory coupled to charged scalar fields have not been carried through, and so the
renormalization group (RG) scalings of several Lorentz-violating couplings in the $SU(2)_{L}$ weak gauge
sector remain undetermined. There are also questions related to higher-loop calculations---such as whether
perturbative renormalizability may be proven to all orders using the same techniques as in conventional
relativistic QFT.

We do not expect that attempts to resolve these kinds of outstanding issues will reveal any
fundamental problems with the minimal SME. For example, power counting arguments that would be very
hard to evade suggest that all sectors of the minimal SME are fully perturbatively renormalizable.
However, these unsolved problems nonetheless present interesting avenues for ongoing and future
research, and the detailed solutions to the problems may reveal new insights into the general structure
of QFT.
For instance, symmetries (internal and external) occupy central roles in
our understanding of the renormalization of the
usual Lorentz-invariant, CPT-invariant standard model. However, these roles will necessarily be subject to
some changes in the context of the SME, since the SME obviously
does not have the same symmetry structure as the unmodified standard model.

Studies of the interplay between renormalization and symmetry in the SME have
already yielded fundamental insights into the nature of finite radiative corrections in
QFT~\cite{ref-jackiw3}.
Among the terms in the quantum electrodynamics (QED) sector of the
minimal SME Lagrange density, the Lorentz-violating Chern-Simons
term is extremely peculiar; the term depends directly on the vector potential $A$ (not just on the
electromagnetic field strength $F$), so that it is not gauge invariant as a density---although
the integrated action is completely gauge invariant.
The peculiar structure of the Chern-Simons term means that many of the usual symmetry arguments
that may normally be applied to the evaluation of radiative corrections are inoperative
in the Lorentz-violating setting.
This provoked quite a bit of controversy over whether
there could ever be a purely radiatively generated Chern-Simons term in the theory and, if so, what
its size should be. Without Lorentz and gauge symmetries to constrain the structure of loop corrections, it
was discovered that with different high-momentum regulators applied to superficially divergent
loop integrals, the theory could produce different finite radiative
corrections~\cite{ref-coleman,ref-jackiw1,ref-victoria1,ref-chung1,
ref-chung2,ref-chen,
ref-chung3,ref-volovik,ref-chan,ref-bonneau1,
ref-chaichian,ref-victoria2,ref-battistel,ref-andrianov,
ref-altschul1,ref-altschul2}. It was subsequently
thought that a nonperturbative approach might be capable of resolving
this confusing situation; however, any nonperturbative regulation procedure that could lead to a nonvanishing
radiative correction term at leading order had to produce an unphysical Lorentz-violating mass-like term in the
photon sector at higher orders.

In this paper, we shall be concerned with another, entirely separate puzzle that appears in the course of
calculating the radiative corrections to the minimal SME. Different arguments,
which appear to be equally valid---a direct one based on
the evaluation of specific Feynman diagrams, and an indirect argument based on known facts about
transformation properties of the SME---seem to give different results for the RG
$\beta$-functions for certain Lorentz-violating operators in the action.

Our treatment is organized as follows. In section~\ref{sec-puzzle}, we introduce the relevant portion
of the SME and lay out an indirect argument for why the RG $\beta$-functions for certain
couplings should be related. In
section~\ref{sec-Of}, we demonstrate, however, that direct calculations of $\beta$-functions do not seem
to bear out the expected relationship. We extend the Feynman diagram calculations to higher orders in the
small SME
parameters in section~\ref{sec-Off}, and in section~\ref{sec-renorm}, we demonstrate how to resolve the
conflict. The
key result is that the $\beta$-functions may depend on the renormalization scheme; this kind of dependence is
common in higher-loop quantum corrections, but here it exists already at one-loop order. Some of these results
are extended to even higher order in the Lorentz violation in section~\ref{sec-Offf}, and
finally, section~\ref{sec-concl} presents an outline of our conclusions.

\section{The $\beta$-Function Puzzle for the SME $f$}
\label{sec-puzzle}

In this paper, we are trying to address a puzzling observation that has been made about the behavior of
quantum corrections in the SME.
The puzzle concerns radiative corrections that arise in the presence of two different kinds of terms
in the SME fermion sector. The Lagrange density for a fermion species in the minimal SME is
\begin{equation}
{\cal L}_{\psi}=\bar{\psi}(i\Gamma^{\mu}\partial_{\mu}-M)\psi,
\end{equation}
where $\Gamma$ and $M$ can include terms with all possible Dirac matrix structures. However, the puzzle
we are interested in concerns only a few of the possible terms, and we may limit attention to
theories with $M=m$ and
\begin{equation}
\Gamma^{\mu}\equiv\gamma^{\mu}+\Gamma_{1}^{\mu}=\gamma^{\mu}+c^{\nu\mu}\gamma_{\nu}+if^{\mu}\gamma_{5}.
\end{equation}
The quantities $c$ and $f$ represent fixed tensor and axial-vector backgrounds that distinguish
physically between different spacetime directions.
In most cases, whether a local Lorentz-violating operator (in the fermion sector or elsewhere)
violates CPT symmetry is determined by whether it has an odd or even
number of outstanding Lorentz indices. A CPT-odd term will ordinarily have an odd number of indices, and
conversely. However, we shall see below that this correspondence
does not hold in quite the way we might expect for the
$f$ term that lies at the center of our puzzle.

The fermionic action must be supplemented with a bosonic
propagation action and a fermion-boson vertex in order to
have nontrivial interactions, including radiative corrections. The puzzle arises whether the quanta
in the boson sector are vector or scalar. In the former case, the QED sector of the minimal SME
has a Lagrange density
\begin{equation}
\label{eq-LA}
{\cal L}_{A}=-\frac{1}{4}F^{\mu\nu}F_{\mu\nu}
-\frac{1}{4}k_{F}^{\mu\nu\rho\sigma}F_{\mu\nu}F_{\rho\sigma}
-e\bar{\psi}\Gamma^{\mu}\psi A_{\mu}.
\end{equation}
The presence of the full Lorentz-violating $\Gamma$ (as opposed to the usual $\gamma$) in the interaction
term is required by gauge invariance. However, (\ref{eq-LA}) is not otherwise
the most general Lorentz-violating Lagrange density for the photon sector, even
in just the minimal SME. The extremely interesting Chern-Simons term mentioned previously has been
omitted, since its discrete symmetries preclude its being involved in the resolution of the puzzle
with $f$ and $c$.
Moreover, only a $k_{F}$ taking the form
\begin{equation}
k_{F}^{\mu\nu\rho\sigma}=\frac{1}{2}\left(g^{\mu\rho}k_{F\alpha}\,^{\nu\alpha\sigma}
-g^{\mu\sigma}k_{F\alpha}\,^{\nu\alpha\rho}
-g^{\nu\rho}k_{F\alpha}\,^{\mu\alpha\sigma}
+g^{\nu\sigma}k_{F\alpha}\,^{\mu\alpha\rho}\right)
\end{equation}
would
need to be considered, for similar reasons. Note that, not entirely coincidentally, the terms
that are included in our ${\cal L}_{A}$ are exactly those minimal SME photon terms that do not lead
to any birefringence for electromagnetic waves propagating in vacuum.

The QED sector of the SME has gotten more attention in the past;
however, for the present work, it probably makes more sense to look at a Yukawa theory, with
bosonic Lagrange density
\begin{equation}
\label{eq-Lphi}
{\cal L}_{\phi}=\frac{1}{2}(\partial^{\mu}
\phi)(\partial_{\mu}\phi)+\frac{1}{2}K^{\mu\nu}(\partial_{\nu}\phi)
(\partial_{\mu}\phi)-\frac{1}{2}\mu^{2}\phi^{2}-\frac{\lambda}{4!}\phi^{4}
-g\bar{\psi}\psi\phi.
\end{equation}
The renormalization of a fully general SME Yukawa theory is more complicated than the analogous
renormalization problem for the SME QED sector. The reason is that the most general theory also
includes additional Yukawa-like interactions, with nontrivial Dirac matrices sandwiched
between the fermion operators in the $\bar{\psi}\cdot\psi\phi$ term.
(Beyond one-loop order, the presence of the four-$\phi$ coupling $\lambda$ also adds
another complication.) However, once again,
none of those nonstandard fermion-boson coupling terms can play a role in our puzzle, and they
have been omitted from ${\cal L}_{\phi}$.

Then Feynman diagram calculations in the Yukawa theory---when we specifically limit
consideration to diagrams with only the 
normal Yukawa coupling $g$---are actually simpler than the analogous calculations in the QED sector.
The reason
now is that the Lorentz-violating $c$ and $f$ in the Yukawa theory only appear as
insertions into the fermion propagator in the scalar theory. In contrast, in the gauge theory
$c$ and $f$ appear in both the propagator and the vertex, and including diagrams with Lorentz-violating
vertices in matrix element calculations will significantly increase their complexity.
For this reason---and because
there does not appear to be any reason to expect any conceptual differences in how the
$\beta$-function puzzle plays out in the two theories---we shall perform all our explicit one-loop
diagram computations in the Yukawa theory.

Among the results obtained in the course of studying the renormalization of the minimal SME were
the RG $\beta$-functions for the Lorentz-violating operators in the two specific theories outlined
above. In both of these theories, it was found that the $\beta$-function for the fermion $f$ vanished
at one loop and linear order in the Lorentz violation itself. At the same time, the $\beta$-functions
for the $c$ terms were generically nonzero.
On one hand, there is nothing
seemingly surprising about the $\beta_{f}$ result. It was observed fairly early on
that there were very frequently no physical consequences to having a $f$ term in the fermion sector
(at least at first order in $f$). The reason seemed straightforward; there were generally no
operators representing physical observables
in the theory that had the right structure to be sensitive to the $f$ coefficient. Note that the operator
corresponding to $f_{0}$ (that is, $i\bar{\psi}\gamma_{5}\partial_{0}\psi$)
is not merely odd under C, P, and T
separately, but it is actually odd under any reflection, regardless of orientation. This is quite different
from the behavior of most P-odd operators---such as $\bar{\psi}\gamma_{j}\psi$, which is odd under a reflection
R$_{j}$
of the $x_{j}$-axis but even under reflections along the the two other perpendicular directions. The spacelike
operators $i\bar{\psi}\gamma_{5}\partial_{j}\psi$ likewise have discrete symmetries (under the full set of
inversions C, T, R$_{1}$, R$_{2}$, and R$_{3}$) that do not match those of any other
normally-available observable.

In fact, the
observation that $if^{\mu}\bar{\psi}\gamma_{5}\partial_{\mu}\psi$ has no observable effects at linear
order
is actually related to another remarkable property of the theory. With a redefinition of the fermion field,
it is actually possible to rewrite the Lagrange density with the $f$ term as one with a $c$ term
instead~\cite{ref-altschul8}. In
other words, the $f$ term is really a $c$ term, combined with a change in the representation of the Dirac
matrices! A transformation 
\begin{equation}
\label{eq-redefgeneral}
\psi'=e^{\frac{i}{2}f^{\mu}\gamma_{\mu}\gamma_{5}G\left(-f^{2}\right)}\psi,
\end{equation}
where $G(\xi)=\frac{1}{\sqrt{\xi}}\tan^{-1}\sqrt{\xi}$ is an analytic function of $\xi=-f^{2}$,
(along with a corresponding transformation
for $\bar{\psi}'$) transforms a Lagrange density for the field $\psi$ with
a $f$ term into one for $\psi'$ with no $f$ term but instead a $c$ term taking the form
\begin{equation}
\label{eq-c}
c^{\nu\mu}=\frac{f^{\nu}f^{\mu}}{f^{2}}\left(\sqrt{1-f^{2}}-1\right)
\approx-\frac{1}{2}f^{\mu}f^{\nu}.
\end{equation}
The approximate form on the right-hand side of (\ref{eq-c}) is valid when all the components of $f$ are small,
but the full expression with the radical
is an exact result, and the transformation is permitted for all $f^{2}<1$ (that is, all $f$ that
do not change the signature of the bilinear form that couples the two factors of the
four-momentum in the fermion dispersion relation).

Thus, there should be
no phenomena that are specific to the presence of a $f$ term. In fact, this could
actually be inferred just from the energy-momentum relations for fermions in the presence of either
a $c$ term or a $f$ term. The two dispersion relations are
\begin{equation}
\label{eq-disprel-c}
(g^{\mu\nu}+c^{\nu\mu}+c^{\mu\nu}+c^{\alpha\mu}c_{\alpha}\,^{\nu})p_{\mu}p_{\nu}-m^{2}=0
\end{equation}
for $c$; and for $f$, 
\begin{equation}
\label{eq-disprel-f}
(g^{\mu\nu}-f^{\mu}f^{\nu})p_{\mu}p_{\nu}-m^{2}=0.
\end{equation}
From these dispersion relations, the equivalence
\begin{equation}
\label{eq-cfequiv}
c^{\nu\mu}\approx-\frac{1}{2}f^{\mu}f^{\nu}
\end{equation}
for small $f$ is once again clearly evident. Note another interesting feature of the $f$ theory that
is illuminated by this equivalence. While the (single-index) $f$ term in the Lagrangian superficially appears
to be odd under CPT, it does not actually give rise to any CPT-odd effects, because it is equivalent to a
two-index, CPT-even $c$ term at ${\cal O}(f^{2})$.

Note also that both dispersion relations, with $c$ and $f$, are quadratic in the energy-momentum $p$. Of course,
the ordinary dispersion relation derived from the Lorentz-invariant Dirac equation, $p^{\mu}p_{\mu}-m^{2}=0$,
is quadratic. However, in the presence of more a general Lorentz-violating kinetic operator
$i\bar{\psi}\Gamma_{1}^{\mu}\partial_{\mu}\psi$
in the fermion action, the dispersion relation---since it is derived
from the determinant of a $4\times 4$ matrix---can be quartic. That the quartic dispersion relation can have
four roots (two positive and two negative for small $\Gamma_{1}$) indicates that the energy can depend on
both the fermion-antifermion identity and spin of an excitation. However, for (\ref{eq-disprel-c}) and
(\ref{eq-disprel-f}) this is not the case; the energies are
both independent of the spin orientation and the same for particles and antiparticles
(thus C even).

It is also evident from (\ref{eq-disprel-c}) that at ${\cal O}(c^{1})$ the energy-momentum relation
depends only on the symmetric part $c^{(\nu\mu)}=c^{\nu\mu}+c^{\mu\nu}$ of the Lorentz-violating background
tensor $c$. In fact, it can be shown, just as with $f$, that there are no physical manifestations of
$c^{[\nu\mu]}=c^{\nu\mu}-c^{\mu\nu}$ at first order. The reasons are actually quite similar. As noted,
at linear order $f$ is equivalent to a change in the representation of the Dirac matrices. Since
$\{\gamma^{\mu},\gamma_{5}\}=0$, the matrices $\Gamma^{\mu}=\gamma^{\mu}+if^{\mu}\gamma_{5}$
in the presence of just a $f$ (no $c$) obey the same
Clifford algebra relations as the $\gamma^{\mu}$,
\begin{equation}
\label{eq-GammaGamma}
\{\Gamma^{\mu},\Gamma^{\nu}\}\approx 2g^{\mu\nu},
\end{equation}
up to corrections that are ${\cal O}(f^{2})$.

In $3+1$ dimensions, there are five mutually anticommuting
Dirac matrices, and at first order, $f^{\mu}$ is just an infinitesimal
rotation of the effective Dirac matrix $\Gamma^{\mu}$
away from the $\gamma^{\mu}$ direction and toward the $\gamma_{5}$ direction.
The field redefinition (\ref{eq-redefgeneral}) just absorbs this
rotation of the Dirac matrices back into the field $\psi'$.
An infinitesimal rotation
of $\Gamma^{\mu}$ toward
a different $\gamma^{\nu}$ (with $\nu\neq\mu$) direction is similarly represented by the inclusion of
$c^{[\nu\mu]}$. With only a $c^{[\nu\mu]}$ present, so that
$\Gamma^{\mu}=\gamma^{\mu}+\frac{1}{2}c_{[\nu}\,^{\mu]}\gamma^{\nu}$,
we again have (\ref{eq-GammaGamma}), up to corrections that are second order in the Lorentz violation
coefficients.
We shall thus henceforth explicitly assume that
$c^{\nu\mu}=c^{\mu\nu}$ is symmetric in its Lorentz indices.
A full understanding of quadratic Lorentz-violating dispersion relations, including the contributions
in (\ref{eq-disprel-c})
that are second order in $c$, is further enabled by making a comparison with
the dispersion relation for the free scalar sector of the ${\cal L}_{\phi}$ from (\ref{eq-Lphi}),
and  this connection can be efficiently explored using supersymmetry~\cite{ref-berger}.

If the $c$ and $f$ terms are actually physically equivalent, then it seems like this ought to show
up in their RG scalings. If the RG $\beta$-function for a quantity $x$ is expressed as $\beta_{x}=x\Psi(x)$,
then it appears that we should have, at leading order, $\Psi(f^{\mu})=\frac{1}{2}\Psi(c^{\nu\rho})$ so that
the RG evolution of $c^{\nu\mu}$ and $f^{\mu}f^{\nu}$ will be equivalent. However, this
has not been borne out
by the explicit calculations of the $\beta$-functions; in particular, $\beta_{f}$
has been found to be vanishing
with either a QED or Yukawa coupling, while $\beta_{c}$ was not.
Obviously, the problem has something to do with the fact that the
calculation of  $\beta_{f}$ only considered terms of first order in $f$, but it is extremely puzzling that
there is this discrepancy between two different ways of finding the function $\beta_{f}$.

The resolution must lie with a fact about QFT that has long been known, but which has not previously been
applied to a situation quite like this one. In actuality, the $\beta$-functions (and other RG functions)
for a theory
are not observable objects unto themselves. A $\beta$-function can depend on
the renormalization conditions
used to define the theory, and it may also potentially depend on the gauge in a gauge theory.
Ordinarily, this issue does not crop up at one-loop order (and often not even at two-loop
order)~\cite{ref-stevenson,ref-chishtie1,ref-ryttov1,ref-chishtie2,ref-mckeon,ref-ryttov2}.
To get a one-loop $\beta$-function, it is typically sufficient just to take a linear combination of the
coefficients of the logarithmically divergent diagrams; this results in an expression that is independent
of quantities like the renormalization scale. While that same approach does not appear to be
{\em wrong} here, neither is the $\beta$-function the approach produces
 uniquely correct and unambiguous. The exact equivalence between
fermion theories with $f$ or $c$ terms marks an ambiguity in how the theory may be represented,
and that ambiguity evidently carries over to the $\beta$-functions, even at just one-loop order.

The fact that there is a degree of reparameterization invariance makes this situation qualitatively
similar to that seen in certain generalized gauge theories. There is a family of
physically equivalent of actions, the parameter in question being the relative sizes of the
equivalent $f$ and $c$ terms.  The parameter may be changed, without affecting the physical
observables of the theory, by a rotation of the basis vectors in the Dirac Grassmann algebra.
When a reparameterization symmetry of the action exists at the classical level, there are often
interesting and nontrivial complications in the lifting of that symmetry to the classical level,
and it becomes a question whether the relations implied by a classical symmetry are stable under
radiative corrections. In the canonical formulation, these complications arise from operator ordering
issues, while in the path integral formalism, the new terms can appear out of the transformed integral
measure~\cite{ref-gervais,ref-christ1}.

It could be interesting to study the consequences of the $f$-$c$ duality in this context, but the
details are beyond the scope of this paper.  The main relevant question would be whether or not
the relation (\ref{eq-cfequiv}) is radiatively stable, provided we begin with a theory which only had
a $f$ term.
Approaching this through the path integral, it appears that there should be no additional terms appearing
in the action when we make the transformation~(\ref{eq-redefgeneral}), and there appear to be no anomaly-like
term from the Jacobian. The full technical details of this analysis are potentially interesting, but
they would take us too far afield from the main thrust of our analysis in this paper. Here we are
focusing specifically on the renormalization group $\beta$-functions, because of the particularly
prominent part that the $\beta$-functions play in the modern understanding of quantum corrections,
and because our results are actually contrary to some general expectations about how $\beta$-functions
ought to behave.

\section{Fermion Self-Energy Up to ${\cal O}(f^{1})$}
\label{sec-Of}

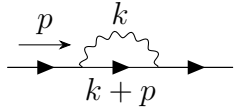
\begin{figure}
\begin{center}
\feynmandiagram [layered layout, horizontal=a to d] {
  a -- b [] -- [fermion, edge label'=\(k+p\)] c [fermion] -- d [],
  c -- [photon, half right, looseness=1.5, edge label'=\(k\)] b,
  a -- [fermion, momentum=\(p\)] b,
  c -- [fermion] d};
\end{center}
\caption{Fermion self-energy diagram, with the virtual emission and reabsorption of a boson,
that would appear in the absence of $f$-type or $c$-type Lorentz violation.
\label{fig-Sigma0}}
\end{figure}

The previously existing
calculations of the fermion self-energy (which led to the conclusion that the one-loop
$\beta$-function for $f$ should vanish) are fairly straightforward. The determination of the $\beta$-function
by the usual method requires the evaluation of two fermion self-energy diagrams. The simplest one is the
usual self-energy diagram in the scalar Yukawa theory without Lorentz violation, as shown in
figure~\ref{fig-Sigma0}. To evaluate divergent diagrams, we shall use dimensional regularization.
(For purposes of extracting the infinite parts of logarithmically divergent loop integrals,
we do not need to introduce a
dimensional extension of the $\gamma_{5}$ anticommutation relation $\{\gamma_{5},\gamma^{\mu}\}=0$. See
Ref.~\cite{ref-altschul2} for further remarks on the extension of $\gamma_{5}$ to nonintegral dimensions
in Lorentz-violating theories.)
The value of the Lorentz-invariant self-energy diagram, in $d=4-\epsilon$ dimensions, is
\begin{eqnarray}
-i\Sigma_{0} & = & (-ig)^{2}\int\frac{d^{d}k}{(2\pi)^{d}}\frac{i(\slashed{k}+\slashed{p}+m)}{(k+p)^{2}-m^{2}}
\frac{i}{k^{2}-\mu^2} \\
& = & g^2\int_{0}^{1}dx\int\frac{d^{d}l}{(2\pi)^{d}}\frac{\slashed{l}+(1-x)\slashed{p}+m}{(l^{2}-\Delta)^{2}},
\end{eqnarray}
using Feynman parameters and $l=k+xp$, so that $\Delta=xm^{2}+(1-x)\mu^{2}-x(1-x)p^{2}$, as usual.
[Conveniently, the
insertion of additional perturbative Lorentz-violating vertices will not change the general
$(l^{2}-\Delta)^{n}$ structure of the higher-order denominators.] The evaluation of the remaining
expression is standard, yielding the logarithmically divergent
\begin{equation}
-i\Sigma_{0}=ig^{2}\int_{0}^{1}dx\,\left[(1-x)\slashed{p}+m\right]\eta;
\end{equation}
the divergence is encapsulated within
\begin{equation}
\eta=\eta(\Delta)=\frac{\Gamma(\epsilon/2)}{(4\pi)^{2-\epsilon/2}\Delta^{\epsilon/2}}.
\end{equation}

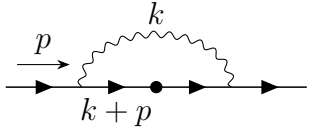
\begin{figure}
\begin{center}
\feynmandiagram [layered layout, horizontal=a to d] {
a -- b [] --e[dot] -- c [] --d[],
b -- [photon, half left, looseness=1.2, edge label=\(k\)] c,
a -- [fermion, momentum=\(p\)] b,
b -- [fermion, edge label'=\(k+p\)] e,
e -- [fermion] c,
c -- [fermion] d,};
\end{center}
\caption{One-loop self-energy diagram, with a single insertion of $f$ (represented by the dot)
along the fermion propagator.
\label{fig-Sigmaf}}
\end{figure}

Along with the conventional self-energy diagram, it is also necessary to evaluate the
amputated one-loop diagram
with the same Lorentz-violating structure as $f$ itself. This is a diagram with a single insertion of
the CPT-odd $f$-dependent vertex $-\gamma_{5}f^{\mu}p_{\mu}$. (Full Feynman rules for the Lorentz-violating
Yukawa theory, including both $f$ and $c$ vertices, are given in~\cite{ref-ferrero3}.)
The diagram, with the Lorentz-violating insertion
represented by a dot, is shown in figure~\ref{fig-Sigmaf}. Its value is
\begin{eqnarray}
\label{eq-Sigmaf-int}
-i\Sigma_{f} & = & (-ig)^{2}\int\frac{d^{d}k}{(2\pi)^{d}}\frac{i(\slashed{k}+\slashed{p}+m)}{(k+p)^{2}-m^{2}}
\left[-\gamma_{5}f^{\mu}(k_{\mu}+p_{\mu})\right]\frac{i(\slashed{k}+\slashed{p}+m)}{(k+p)^{2}-m^{2}}
\frac{i}{k^{2}-\mu^2} \\
& = & -g^{2}\gamma_{5}f^{\mu}p_{\mu}\int_{0}^{1}dx\,(1-x)\eta.
\end{eqnarray}
The integrand of (\ref{eq-Sigmaf-int}) was simplified by using $(\slashed{k}+\slashed{p}+m)\gamma_{5}
(\slashed{k}+\slashed{p}+m)=-\gamma_{5}[(k+p)^{2}-m^{2}]$. Then,
as already noted, the use of Feynman parameters to simplify the denominator is unchanged from
$\Sigma_{0}$, since the fermion propagators situated on either side of the $f$ vertex have exactly the same
momenta. This is responsible for the simplification of the magnitude of the infinite part of $\Sigma_{f}$
to a form very similar to that previously seen in $\Sigma_{0}$.

Extracting the divergence in the self-energy, we have
\begin{equation}
\label{eq-Sigmaf-inf}
-i\Sigma_{f}\overset{{\rm LV}}{\sim}-\frac{g^{2}}{2}\left(\gamma_{5}f^{\mu}p_{\mu}\right)\eta.
\end{equation} 
The $\overset{{\rm LV}}{\sim}$ notation in (\ref{eq-Sigmaf-inf}) indicates that the two expressions have the
same Lorentz-violating divergent parts. To isolate the divergence we evaluate $\eta$ at a
renormalization scale $\Delta=M^{2}$, as usual.

Then the fact that the infinite coefficients of
$i\gamma_{5}f^{\mu}p_{\mu}$ and $\slashed{p}$ in the two expressions are the same is what produces the
vanishing $\beta$-function in this renormalization scheme. Following the usual procedure for renormalization,
there must be a fermion field strength renormalization counterterm, $\delta_{\psi}=Z_{\psi}-1$ given by
the infinite part of the $\slashed{p}$ coefficient in $\Sigma_{0}$, renormalized at $M^{2}$,
\begin{equation}
\delta_{\psi}=-\frac{g^{2}}{2}\eta(M^{2}).
\end{equation}
There is also a counterterm $\delta_{f}^{\mu}=(Z_{f}-1)f^{\mu}$,
\begin{equation}
\label{eq-deltaf1}
\delta_{f}^{\mu}=-\frac{g^{2}}{2}f^{\mu}\eta(M^{2}).
\end{equation}
Following the usual procedure of just reading off the coefficients of
$\Gamma(\epsilon/2)/(M^{2})^{\epsilon/2}\equiv\log(\Lambda^{2}/M^{2})$ (reexpressing the dimensional
cutoff $\epsilon$ in terms of an effective energy-momen\-tum cutoff scale $\Lambda$),
\begin{equation}
\beta_{f}^{\mu}=\beta_{f^{\mu}}=M\frac{\partial}{\partial M}\left(-\delta_{f}^{\mu}+f^{\mu}\delta_{\psi}
\right)=0.
\end{equation}
This seems naively like an unambiguous result, but the true situation is actually more subtle, and it is
possible to set physically motivated renormalization conditions in a different way, so as to obtain an
entirely different answer!

\section{Self-Energy at ${\cal O}(f^{2})$ and ${\cal O}(c^{1})$}
\label{sec-Off}

The renormalization conditions used in section~\ref{sec-Of} were so standard that we did not even
spell them out explicitly. In essence, the renormalization condition that set $\delta_{f}$ was one
that forced the tree-level plus one-loop contributions to the $f$ term in the fermion propagator to take
a certain value. Normally, in the discussion of some other parameter (such as a coupling constant like $g$,
or another SME parameter like $c$), we might say that those particular renormalization conditions
forced the parameter in question to take its ``physical'' value. However, that is not possible with $f$,
because there is no ``physical'' value of $f$ at the order we have so far considered. Remember, there is
no physical observable in our theory that differs from its value in the Lorentz-symmetric theory at
${\cal O}(f^{1})$; nothing we can measure in the theory depends linearly on $f$.

The next natural question is how the renormalization conditions may be set in order to
guarantee that we instead have $\Psi(f^{\mu})=\frac{1}{2}\Psi(c^{\nu\rho})$. In fact, there is a continuum of
possible renormalization frameworks for the theory. The source of the ambiguity is precisely the fact
that nothing physically observable depends on the
value of $\delta_{f}$ at ${\cal O}(f^{1})$. The presence of a $\delta_{f}$ counterterm in the Feynman rules
does not
lead to any physically meaningful changes in a theory, unless it appears in a diagram in conjunction with
at least one more factor of $f$. This is just a consequence of the form taken by the counterterm; although the 
counterterm contains a formal infinity, it has the same Lorentz structure
as a bare $f$ term in the Lagrange density---which we know is not observable on its own. What this ultimately
means is that the value $\delta_{f}$ is not actually uniquely determined (or even at all restricted)
by the structure of any ${\cal O}(f^{1})$ loop diagrams. The one-loop counterterm may instead be chosen so as to
absorb the physical infinities that arise from a diagram with not one, but two, $f$ vertex insertions.

So in this section we shall evaluate a fermion self-energy diagram in which there are two $f$ insertions along
the internal fermion line. We shall also look at the diagram
with a single $c$ insertion, since---under generic
renormalization conditions---the $f$ and $c$ parameters will mix under radiative corrections, with the
lowest-order $c$ that may be generated purely from $f$ being of ${\cal O}(f^{2})$.
(The full evaluation of the fermion self-energy at this order should really also include
the evaluation of a diagram with a $K$ insertion on the boson propagator. However, while $c$ and $K$
do mix under the action of the RG, the $K$-dependent radiative corrections do not play any essential role
in the resolution of the $\beta$-function puzzle. We shall therefore not consider them any further.)

It is not, of course, unexpected that radiative corrections at ${\cal O}(f^{2})$ can give rise to a $c$-type
term in the fermion self-energy. The product $f^{\nu}f^{\mu}$ has exactly the right discrete symmetries
to generate an effective $c^{\nu\mu}$. In general, beyond first order, the SME coefficients may mix in
increasingly complicated ways~\cite{ref-brito1}. What is novel to this discussion is the observation that
there is actually a freedom to assign certain radiative corrections to be renormalizations of either $c$
or $f$.

The equivalence between the insertion of two $f$ vertices and a single $c$ vertex on an on-shell
fermion line is easy
to see. A fermion line carrying momentum $p$ with three propagators and two $f$ insertions takes the form
\begin{eqnarray}
\label{eq-SfSfS}
S(p)(-\gamma_{5}f^{\nu}p_{\nu})S(p)(-\gamma_{5}f^{\mu}p_{\mu})S(p)
& = & \frac{i(\slashed{p}+m)}{p^{2}-m^{2}}(\gamma_{5}f^{\nu}p_{\nu})\frac{i(\slashed{p}+m)}{p^{2}-m^{2}}
(\gamma_{5}f^{\mu}p_{\mu})\frac{i(\slashed{p}+m)}{p^{2}-m^{2}} \quad\quad \\
\label{eq-ff}
& = & (-i)\frac{(\slashed{p}+m)}{p^{2}-m^{2}}(f^{\nu}f^{\mu}p_{\mu}p_{\nu})
\frac{(-\slashed{p}+m)(\slashed{p}+m)}{(p^{2}-m^{2})^{2}} \\
\label{eq-ffred}
& = & \frac{i(\slashed{p}+m)}{(p^{2}-m^{2})^{2}}(f^{\nu}f^{\mu}p_{\mu}p_{\nu}),
\end{eqnarray}
moving a $\gamma_{5}$ past the middle propagator in order to cancel it out in (\ref{eq-ff}).
Alternatively,
since $(\slashed{p}-m)$ and $(\slashed{p}+m)$ commute, we may write the overall numerator of (\ref{eq-ff})
as $if^{\nu}f^{\mu}p_{\mu}(\slashed{p}+m)p_{\nu}(\slashed{p}+m)(\slashed{p}-m)$.
When $p$ is on shell, by further invoking the
closure identity for Dirac spinors,
\begin{equation}
\slashed{p}+m=\sum_{s}u_{s}(p)\bar{u}_{s}(p),
\end{equation}
we may sandwich $p_{\nu}$ between momentum eigenspinors. Then using the Gordon identity at
zero momentum transfer,
\begin{equation}
\bar{u}(p)p_{\nu}u(p)=\bar{u}(p)(m\gamma_{\nu})u(p),
\end{equation}
and the Dirac eigenvalue condition for the spacetime-independent spinor $u(p)$, which is $(\slashed{p}-m)u(p)=0$,
we may rewrite part of the numerator from (\ref{eq-ff}) with
\begin{equation}
(\slashed{p}+m)p_{\nu}(\slashed{p}+m)=\frac{1}{2}(\slashed{p}+m)\gamma_{\nu}(\slashed{p}+m)^{2}.
\end{equation}
Returning to the full expression (\ref{eq-ff}), including the denominators, we have, for on-shell $p$,
\begin{eqnarray}
\label{eq-SffS}
S(p)(-\gamma_{5}f^{\nu}p_{\nu})S(p)(-\gamma_{5}f^{\mu}p_{\mu})S(p)=
S(p)\left(-\frac{i}{2}f^{\nu}f^{\mu}\gamma_{\nu}p_{\mu}\right)S(p).
\end{eqnarray}
This has exactly the form of a $c$ insertion into a fermion line, and the preceding calculation 
is actually another way of demonstrating an exact equivalence between a theory with a fermion propagation
Lagrangian containing a $c$ term and one containing a $f$ term. The tree-level fermion two-point function for a
fermion field with a $f$ term will involve a sum of diagrams with all possible numbers of $f$ insertions along
the propagator line. According to (\ref{eq-SffS}),
the resummation of all the diagrams with even numbers of $f$ vertices will proceed in
exactly the same way as the resummation of terms with various numbers of $c$ insertions in a fermion theory with
a $c$ coefficient. (The terms with odd numbers of $f$ insertions are, on the other hand, never
directly observable.)
The coefficient of the
middle term in parentheses in (\ref{eq-SffS}) also matches (\ref{eq-cfequiv}), although one
might conceivably
wonder why then (\ref{eq-cfequiv}) is not exact, rather than an ${\cal O}(f^{2})$ approximation. The
reason for this last apparent discrepancy is actually that the presence of a $c$ or $f$ term in the action
affects the canonical normalization of the fermion field at higher orders; the higher-order corrections in
(\ref{eq-c}) are correspondingly only needed to correct for these normalization differences.

\begin{figure}
\begin{center}
\feynmandiagram [layered layout, horizontal=a to d] {
a -- b []--e[dot]-- [fermion,edge label'=\(k+p\)] g[dot]-- c [] --d[],
b -- [photon, half left, looseness=1.0, edge label=\(k\)] c,
a -- [momentum=\(p\)] b,
a -- [fermion] b,
b -- [fermion] e,
g -- [fermion] c,
c -- [fermion] d,};
\end{center}
\caption{Self-energy diagram with two $f$ vertex insertions on the internal fermion line.
\label{fig-Sigmaff}}
\end{figure}
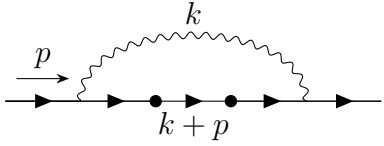

The preceding equivalence was handled entirely at the classical level, but it will provide some
useful illumination for our evaluation of the ${\cal O}(f^{2})$ and higher loop corrections. 
In particular, we shall apply the results (\ref{eq-ff}--\ref{eq-ffred})
to help us evaluate counterterm diagrams that include
$f^{\mu}\delta^{\nu}_{f}$, which possess a Lorentz structure identical to (\ref{eq-SfSfS}).

The ${\cal O}(f^{2})$ diagram with a single one-particle irreducible loop is shown in figure~\ref{fig-Sigmaff}.
Its value is
\begin{eqnarray}
-i\Sigma_{ff} & = & (-ig)^{2}\int\frac{d^{d}k}{(2\pi)^{d}}\frac{i(\slashed{k}+\slashed{p}+m)}{(k+p)^{2}-m^{2}}
\left[-\gamma_{5}f^{\mu}(k_{\mu}+p_{\mu})\right]\frac{i(\slashed{k}+\slashed{p}+m)}{(k+p)^{2}-m^{2}} \nonumber\\
& & \times\left[-\gamma_{5}f^{\mu}(k_{\mu}+p_{\mu})\right]\frac{i(\slashed{k}+\slashed{p}+m)}{(k+p)^{2}-m^{2}}
\frac{i}{k^{2}-\mu^2} \\
& = & g^{2}\int\frac{d^{d}k}{(2\pi)^{d}}\frac{f^{\mu}f^{\nu}(\slashed{k}+\slashed{p}+m)
(k+p)_{\mu}(k+p)_{\nu}}{[(k+p)^{2}-m^{2}]^2(k^{2}-\mu^{2})} \\
& = & g^{2}f^{\mu}\!f^{\nu}\!\!\int_{0}^{1}\!\!dx\, 2x\!\!\int\!\!\frac{d^{d}l}{(2\pi)^{d}}
\frac{[l+(1-x)p]_{\mu}[l+(1-x)p]_{\nu}[\slashed{l}+(1-x)\slashed{p}+m]}{(l^{2}-\Delta)^{3}}.
\label{eq-ffintegral}
\end{eqnarray}
To extract the divergent part of the self-energy (which is what determines the RG behavior), we
can restrict attention to the terms in the numerator that are logarithmically divergent by power
counting---that
is, the terms quadratic in the shifted integration momentum $l$. This lets us reduce the numerator in the
integrand of (\ref{eq-ffintegral}) to
\begin{equation}
N=l_{\mu}l_{\nu}[(1-x)\slashed{p}+m]+l_{\mu}(1-x)p_{\nu}\slashed{l}+(1-x)p_{\mu}l_{\nu}\slashed{l}.
\label{eq-Nff}
\end{equation}
Inside a symmetric integration, we may make the usual replacement of $l_{\alpha}l_{\beta}$ with
$l^{2}g_{\alpha\beta}/4$.
(Dimensional regularization
corrections to this expression vanish as $\epsilon\rightarrow0$, and so can only contribute to
unimportant finite terms.) Moreover the second and third terms on the right-hand side of (\ref{eq-Nff})
are equal when contracted with $f^{\mu}f^{\nu}$. This leaves a reduced numerator
\begin{equation}
N=\frac{l^{2}}{4} g_{\mu\nu}[(1-x)\slashed{p}+m]+\frac{l^{2}}{2}(1-x)p_{\mu}\gamma_{\nu}.
\end{equation}
The term with $g_{\mu\nu}$ contributes, after contraction with $f^{\mu}f^{\nu}$, only to ${\cal O}(f^{2})$
modifications of the fermion mass and field strength renormalization. The second term, in contrast, has a
structure corresponding to a radiatively generated $c^{\nu\mu}$ term.

So the surviving Lorentz-violating contributions to $N$ can be inserted back into (\ref{eq-ffintegral}) to give
\begin{eqnarray}
\label{eq-Sigmaff-LV}
-i\Sigma_{ff} & \overset{{\rm LV}}{\sim} & g^{2}f^{\mu}f^{\nu}p_{\mu}\gamma_{\nu}\int_{0}^{1}dx\,x(1-x)
\int\frac{d^{d}l}{(2\pi)^{d}}\frac{l^{2}}{(l^{2}-\Delta)^{3}} \\
& \overset{{\rm LV}}{\sim} & i\frac{g^{2}}{6}\left(f^{\mu}f^{\nu}p_{\mu}\gamma_{\nu}\right)\eta.
\end{eqnarray}
The $\epsilon\rightarrow0$
infinity in this radiative correction needs to be canceled through the use of a counterterm, although there
are actually several ways that the cancellation may be achieved, combining $\delta_{f}$ and $\delta_{c}$
counterterms in a potentially intricate way.

However, since it looks as if the ${\cal O}(f^{2})$ contribution
to the fermion self-energy may include
a radiatively-generated
$c$ term, we should also look at the renormalization of the theory with a $c$ in the fermion sector. If the
renormalized theory is to contain an effective $c$ term generated by a logarithmic divergence,
then the action for
the bare theory must already include a $c$, so that the infinite correction can be absorbed. We shall therefore
consider the theory with $c$ (in addition to $f$), up to ${\cal O}(c^{1})$. The diagram we need
to compute is again
the one depicted in figure~\ref{fig-Sigmaf}, except that we now interpret the dot on the fermion line as a
$ic^{\nu\mu}\gamma_{\nu}(k+p)_{\mu}$ insertion. Like the ${\cal O}(f^{1})$ loop diagram, this is equivalent to
a calculation already outlined in~\cite{ref-ferrero3}.

Structurally, the diagram with the single $c$ vertex on the internal fermion line is quite similar to the
$\Sigma_{ff}$ diagram.
[In fact, we almost could have used (\ref{eq-SffS}) directly to convert the propagator with two $f$ insertions
into one with a single $c$-like insertion. However, the derivation of (\ref{eq-SffS}) made assumptions about the
propagation being on the mass shell, which
actually make things a bit more subtle than they might initially appear.]

Evaluating the $c$ diagram directly, we find
\begin{eqnarray}
-i\Sigma_{c} & = & (-ig)^{2}\int\frac{d^{d}k}{(2\pi)^{d}}\frac{i(\slashed{k}+\slashed{p}+m)}{(k+p)^{2}-m^{2}}
\left[ic^{\nu\mu}\gamma_{\nu}(k+p)_{\mu}\right]
\frac{i(\slashed{k}+\slashed{p}+m)}{(k+p)^{2}-m^{2}}\frac{i}{k^{2}-\mu^{2}} \\
& = & -g^{2}\int\frac{d^d k }{(2\pi)^d}\frac{c^{\nu\mu}(\slashed{k}+\slashed{p}+m)\gamma_{\nu}(k+p)_{\mu}
(\slashed{k}+\slashed{p}+m)}{[(k+p)^{2}-m^{2}]^{2}(k^2-\mu^{2})}.
\end{eqnarray}
In this case, the insertion of the Feynman parameter yields
\begin{equation}
-i\Sigma_{c}= -g^{2}c^{\nu\mu}\int_{0}^{1}dx\, 2x\int\frac{d^{d}l}{(2\pi)^{d}}
\frac{[\slashed{l}+(1-x)\slashed{p}+m]\gamma_{\nu}[l+(1-x)p]_{\mu}[\slashed{l}+(1-x)\slashed{p}+m]}
{(l^{2}-\Delta)^{3}}.
\end{equation}
Structurally, the divergent (quadratic in $l$) part of the numerator is 
\begin{equation}
N=\slashed{l}\gamma_{\nu}l_{\mu}[(1-x)\slashed{p}+m]+\slashed{l}\gamma_{\nu}(1-x)p_{\mu}\slashed{l}
+[(1-x)\slashed{p}+m]\gamma_{\nu}l_{\mu}\slashed{l}
\end{equation}
or, equivalently,
\begin{equation}
N=\frac{l^{2}}{4}\left\{
\gamma_{\mu}\gamma_{\nu}[(1-x)\slashed{p}+m]-2\gamma_{\nu}(1-x)p_{\mu}+
[(1-x)\slashed{p}+m]\gamma_{\nu}\gamma_{\mu}\right\},
\end{equation}
using $l_{\alpha}l_{\beta}\rightarrow l^{2}g_{\alpha\beta}/4$ as well as
$g_{\alpha\beta}\gamma^{\alpha}\gamma^{\nu}\gamma^{\beta}=\gamma^{\alpha}\gamma^{\nu}\gamma_{\alpha}
=-2\gamma^{\nu}$. We may also take advantage of the identity
\begin{equation}
\slashed{p}\gamma_{\nu}\gamma_{\mu}=2p_{\nu}\gamma_{\mu}-2\gamma_{\nu}p_{\mu}+\gamma_{\nu}\gamma_{\mu}
\slashed{p};
\label{eq-doublecomm}
\end{equation}
contraction of (\ref{eq-doublecomm})
with the symmetric $c^{\nu\mu}$ cancels the first two terms on the right-hand side. This leaves
\begin{equation}
\label{eq-Nc}
N=\frac{l^{2}}{2}\left\{\gamma_{\nu}\gamma_{\mu}[(1-x)\slashed{p}+m]-\gamma_{\nu}(1-x)p_{\mu})\right\}.
\end{equation}

Once again, the self-energy splits into two terms: one which is Lorentz symmetric and represents a minuscule
$c$-dependent modification of the usual field strength and mass renormalizations; and another which has the
form of a radiatively generated $c$. That the first term of (\ref{eq-Nc}) is Lorentz invariant is another
consequence of the $c^{\nu\mu}=c^{\mu\nu}$
symmetric form, combined with $\gamma_{\nu}\gamma_{\mu}=g_{\mu\nu}+i\sigma_{\mu\nu}$.
As in the final evaluation of the $\Sigma_{ff}$ integral, we have here
\begin{equation}
\label{eq-Sigmac-LV}
-i\Sigma _{c}\overset{{\rm LV}}{\sim}i\frac{g^{2}}{6}\left(c^{\nu\mu}\gamma_{\nu}p_{\mu}\right)\eta.
\end{equation}

\section{Alternative Renormalization Schemes}
\label{sec-renorm}

At this point, it is possible to lay out how divergences like (\ref{eq-Sigmaff-LV}) and (\ref{eq-Sigmac-LV})
may be canceled via counterterms, and how the counterterms involved are not unique. One approach is obvious;
a single counterterm
\begin{equation}
\label{eq-deltac}
\bar{\psi}\left[\delta_{c}^{\nu\mu}\gamma_{\nu}(i\partial_{\mu})\right]\psi=
-\frac{g^{2}}{6}\eta\left(c^{\nu\mu}+f^{\nu}f^{\mu}\right)\bar{\psi}\gamma_{\nu}(i\partial_{\mu})\psi
\end{equation}
serves to cancel both the ${\cal O}(c^{1})$ and ${\cal O}(f^{2})$ divergences.

\begin{figure}
\begin{center}
\begin{subfigure}[b]{0.4\textwidth}
\begin{center}
\feynmandiagram [layered layout, horizontal=a to c] {
a -- b [crossed dot]-- [fermion] g[dot]-- c [],
a -- [momentum=\(p\)] b,
a -- [fermion] b,
g -- [fermion] c,};
\end{center}
\end{subfigure}
\begin{subfigure}[b]{0.4\textwidth}
\begin{center}
\feynmandiagram [layered layout, horizontal=a to c] {
a -- b [dot]-- [fermion] g[crossed dot]-- c [],
a -- [momentum=\(p\)] b,
a -- [fermion] b,
g -- [fermion] c,};
\end{center}
\end{subfigure}
\end{center}
\caption{The two diagrams, incorporating both $f$ (dot) and $\delta_{f}$ (circled cross)
vertices,
that have the correct structure to cancel the Lorentz-violating divergence in $\Sigma_{ff}$.
\label{fig-f-deltaf}}
\end{figure}
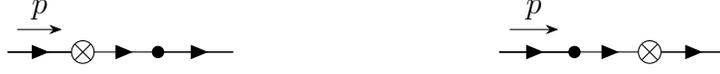

However, it is also possible to cancel the ${\cal O}(f^{2})$ radiative divergences using a $\delta_{f}$
counterterm---remembering that the $\delta_{f}$ was actually not constrained by the value of the self-energy
at ${\cal O}(f^{1})$. There is a bit of subtlety here, because the $\delta_{f}$ on its own clearly has the
wrong spacetime structure to cancel a term like (\ref{eq-Sigmaff-LV}). The presence of both a divergent
$\delta_{f}$ and an additional tree-level $f$ are
actually necessary to effect the cancellation. The diagrams involved
are shown in figure~\ref{fig-f-deltaf}. According to (\ref{eq-SffS}), a
\begin{equation}
\bar{\psi}\left[i\delta_{f}^{\mu}\gamma_{5}(i\partial_{\mu})\right]\psi=
-\frac{g^{2}}{6}\eta\left(if^{\mu}\right)\bar{\psi}\gamma_{5}(i\partial_{\mu})\psi
\end{equation}
counterterm provides an equally plausible way of canceling
the divergence from figure~\ref{fig-Sigmaff}.

To verify the coupling constant flow---whether for $c$ or $f$---we ultimately need to look
at the Callan-Symanzik equation (CSE). For the theory with only a $c$, the CSE for the fermion two-point
correlation function takes the usual form, 
\begin{eqnarray}
\label{eq-CSE}
\left.\left[M\frac{\partial}{\partial M}+\sum_{x_{i}}\beta_{x_{i}}\frac{\partial}{\partial x_{i}}
+2\gamma_{\psi}\right]G^{(2,0)}(\{p_{i}\},\{x_{i}\},M)\right|_{-M^2}=0,
\end{eqnarray}
with two $\beta$-functions: the usual one for the Yukawa coupling $g$ and another one $\beta_{c}^{\nu\mu}$
describing the RG behavior of the the Lorentz-violating $c$ in the action.
Taking the ${\cal O}(\hbar)$ correlation function as being a sum
including [in addition to the usual $G_{0}^{(2,0)}$
from the Lorentz-invariant theory] the one-loop self-energy diagram with the internal leg $c$ insertion and the
counterterm diagram containing $\delta_{c}$ leads immediately to
\begin{equation}
\beta_{c}^{\nu\mu}=M\frac{\partial}{\partial M}\left(-\delta_{c}^{\nu\mu}+c^{\nu\mu}\delta_{\psi}
\right).
\end{equation}
Using the value of $\delta_{c}$ read off from (\ref{eq-deltac}) for the case of the theory with
$c$-type Lorentz violation only,
\begin{equation}
\beta_{c}^{\nu\mu}=\frac{2g^{2}}{3(4\pi)^{2}}c^{\nu\mu}.
\end{equation}
This agrees with the result found in~\cite{ref-ferrero3}, in spite of the rather different bookkeeping
for divergences used in that paper.

If we posit a theory in which the Yukawa coupling $g$ and
the Lorentz-violating $f$ are the only couplings, then the same general form for the CSE (\ref{eq-CSE}) still
applies. However, when we include the $f$-dependent diagrams from figures~\ref{fig-Sigmaff}
and~\ref{fig-f-deltaf}, we find that solving for $\beta_{f}$ actually gives
\begin{eqnarray}
\label{eq-betafsolve}
2\beta_{f}^{\mu} & = & M\frac{\partial}{\partial M}\left(-\delta_{f}^{\mu}+f^{\mu}\delta_{\psi}\right) \\
\label{eq-betaf}
\beta_{f}^{\mu} & = & \frac{g^{2}}{3(4\pi)^{2}}f^{\mu}.
\end{eqnarray}
The key element is the factor of two on the left-hand side of (\ref{eq-betafsolve}). The factor comes
about because there are the two diagrams containing $\delta_{f}$ shown in figure~\ref{fig-f-deltaf}
%(each of which also has both external legs and an internal propagator affected by the field strength
%renormalization)
that both contribute to the two-point correlation function.
It is really as if we were calculating a $\beta$-function for the
power $f^{2}$, rather than $f$ itself. The
mathematical effect is seemingly to spread the RG flow across the two powers of $f$, so that the rate of
RG flow for the physically observable tensor quantity $f^{\nu}f^{\mu}$ is the same as the flow for
the $c^{\nu\mu}$ in an equivalent theory.
Because there are two $\delta_{f}$ diagrams that contribute, the scaling
coefficient of the $\delta_{f}$ needed to cancel the ${\cal O}(f^{2})$ divergence is half the
$\delta_{c}$ required for the cancellation. This is what we argued for above on physical grounds.

[It is actually possible to include the diagram from figure~\ref{fig-Sigmaf} as a piece of a
pair of (one-particle reducible) diagrams at ${\cal O}(f^{2})$ with additional $f$ insertions along
the external legs, without changing the $\beta$-function result (\ref{eq-betaf}). Including these
diagrams just adds and subtracts extra terms looking like (\ref{eq-deltaf1}) in various places, without
changing the $\beta$-functions.]

However, with the general structure of all the (one-loop) ${\cal O}(f^{2})$ and ${\cal O}(c^{1})$ radiative
corrections worked out, we are actually in a position to make a more general statement. In a theory with
(the possibility of) both $c$ and $f$, we actually have a continuous family of choices for how to handle
the counterterms. The divergences will be adequately canceled by any combination of counterterms
\begin{eqnarray}
\delta_{c}^{\nu\mu} & = & -\frac{g^{2}}{6}\eta\left(c^{\nu\mu}+Xf^{\nu}f^{\mu}\right) \\
\delta_{f}^{\mu} & = & -\frac{g^{2}}{6}\eta\left(1+2X\right)f^{\mu}.
\end{eqnarray}
The parameter $X$ can take any real value, with $X=0$ representing the
physically motivated choice we have now discussed extensively. Correspondingly, the RG $\beta$-functions are
\begin{eqnarray}
\label{eq-betacX}
\beta_{c}^{\nu\mu} & = & \frac{g^{2}}{3(4\pi)^{2}}\left(2c^{\nu\mu}-Xf^{\nu}f^{\mu}\right) \\
\label{eq-betafX}
\beta_{f}^{\mu} & = & \frac{g^{2}}{3(4\pi)^{2}}\left(1-X\right)f^{\mu}.
\end{eqnarray}

In a theory with $c$ and $f$ simultaneously present,
those two Lagrangian parameters cannot be measured independently. To
the order we have considered so far, the energy-momentum relation and other
physical quantities only depend on the combination $2c^{\nu\mu}-f^{\nu}f^{\mu}$.
According to
(\ref{eq-betacX}--\ref{eq-betafX}), the RG flow for the physically meaningful combination is described by
\begin{equation}
\label{eq-betacff}
\frac{\partial}{\partial(\log p/M)}\left(c^{\nu\mu}-\frac{1}{2}f^{\nu}f^{\mu}\right)=\frac{2g^{2}}{3(4\pi)^{2}}
\left(c^{\nu\mu}-\frac{1}{2}f^{\nu}f^{\mu}\right),
\end{equation}
which is independent of the renormalization scheme parameter $X$ and
structurally the same as the RG flow in a theory with just a $c$ tensor and no $f$ at all.

What we have uncovered is that the RG functions for the SME with a fermion $f$ term are not unique; they
depend on the particular renormalization scheme. In particular, there are multiple ways to select the
counterterm diagrams that will cancel the divergences that appear in the ${\cal O}(f^{2})$ fermion
self-energy. From one viewpoint, this is actually rather unsurprising. Even at tree level, the description
of the theory contains redundancies; it is possible to exchange a $f$ coefficient for an equivalent $c$ in
any classical perturbative calculation. What these results show is that the ambiguity extends to the
quantum level.

However, on the other hand, explicit scheme dependence is not something that is usually seen in the
one-loop RG structure of perturbatively coupled theories. Explicit scheme dependence typically enters
at two- or three-loop orders.
We have therefore identified another way in which the SME can provide new insights into the general
structure of QFTs. There is also a degree of commonality with previous results, in that
explicit
scheme dependence is frequently associated with situations in which the physical phenomena are distinctly
nonlinear functions of the scale-dependent coupling paramaters; this is also exactly what happens in the
SME with the effects of the $f$ term, which
can only contribute to physically observable effects nonlinearly.

\section{Higher-Order Radiative Corrections}
\label{sec-Offf}

In approaching this problem, we initially thought that finding the alternative renormalization conditions
that would ensure $\Psi(f^{\mu})=\frac{1}{2}\Psi(c^{\nu\rho})$
might be facilitated by extending the one-loop radiative correction calculations to ${\cal O}(f^{3})$ and
${\cal O}(c^{1}f^{1})$. However, after some
further consideration, it became clear that the resolution to the puzzle could not
really involve anything beyond ${\cal O}(f^{2})$ and ${\cal O}(c^{1})$
in a fundamental way. The reason is that
the ${\cal O}(f^{3})$ loop corrections can only produce potentially divergent
corrections to the fermion propagator that involve the
structure $i\gamma_{5}f^{2}f^{\mu}p_{\mu}$. The corrections will have this structure even in the
lightlike, $f^{2}=0$, case, but in that case, those corrections are manifestly vanishing. In this special case
the solution cannot involve imposing conditions on the ${\cal O}(f^{3})$ terms. However, if the method of solution
involves a power-series expansion in the components of $f$, then it should apply just as well at
$f^{2}=0$ as for other values of $f^{2}$.
(In fact, one might actually expect the theory with $f^{2}=0$ to have the most straightforward
behavior. The lightlike case has the simplest and best-behaved
correspondence between the $f$ and an effective $c$, because the $c^{\nu\mu}$ equivalent of a
lightlike $f$ is {\em exactly} $-\frac{1}{2}f^{\nu}f^{\mu}$; higher order corrections are impossible simply
by virtue of $f^{2}$ being zero. Moreover, the quantity $-\frac{1}{2}f^{\nu}f^{\mu}$ is traceless,
and so it is equivalent to a
$c$ that does not disturb the canonical normalization of the fermion field.)

Nevertheless, we believe it is sufficiently interesting to record here the results of the fermion
self-energy calculation at ${\cal O}(f^{3})$ and ${\cal O}(c^{1}f^{1})$.
We consider first the diagram with three $f$ insertions along in the internal fermion line. The calculation for
this
diagram goes in a very similar way to ones we have already done so far. The self-energy is
\begin{eqnarray}
-i\Sigma_{fff} & = & (-ig)^{2}\int\frac{d^{d}k}{(2\pi)^{d}}\frac{i(\slashed{k}+\slashed{p}+m)}{(k+p)^{2}-m^{2}}
\left\{\left[-\gamma_{5}f^{\mu}(k_{\mu}+p_{\mu})\right]\frac{i(\slashed{k}+\slashed{p}+m)}{(k+p)^{2}-m^{2}}
\right\}^{3} \\
& = & ig^{2}\int\frac{d^{d}k}{(2\pi)^{d}}
\frac{\gamma_{5}f^{\mu}f^{\nu}f^{\rho}(k+p)_{\mu}(k+p)_{\nu}(k+p)_{\rho}}{[(k+p)^{2}-m^{2}]^2(k^{2}-\mu^{2})}.
\end{eqnarray}
Performing the $l=k+xp$ substitution again, with the usual algebra, gives 
\begin{equation}
\label{eq-Sigmafff-int}
-i\Sigma_{fff}=ig^{2}\gamma_{5}f^{\mu}f^{\nu}f^{\rho}\int_{0}^{1}dx\,2x\int\frac{d^{d}l}{(2\pi)^{d}}
\frac{[l+(1-x)p]_{\mu}[l+(1-x)p]_{\nu}[l+(1-x)p]_{\rho}}{(l^2-\Delta)^3}.
\end{equation}
Again taking only the quadratic part of the numerator then simplifies the necessary numerator to 
\begin{equation}
\label{eq-Nfff}
N=(1-x)(l_{\mu}l_{\nu}p_{\rho}+l_{\mu}l_{\rho}p_{\nu}+l_{\nu}l_{\rho}p_{\mu}).
\end{equation}
Since the numerator $N$ is contracted with $f^{\mu}f^{\nu}f^{\rho}$, the three terms in the numerator
contribute equally. Hence we obtain the equivalent numerator
\begin{equation}
\label{eq-Nfff-symm}
N=\frac{3}{4}(1-x)l^{2}g_{\mu\nu}p_{\rho},
\end{equation}
so completing the calculation gives the infinite part as 
\begin{eqnarray}
-i\Sigma_{fff} & \overset{{\rm LV}}{\sim} &
-\frac{3g^2}{2}\gamma_{5}f^{\mu}f^{\nu}f^{\rho}g_{\mu\nu}p_{\rho}\int dx\,x(1-x )\,\eta \\
& \overset{{\rm LV}}{\sim} & -\frac{g^{2}}{4}\left(\gamma_{5}f^{2}f^{\mu}p_{\mu}\right)\eta.
\end{eqnarray}
We see that this
indeed has the same Lorentz structure as the term (\ref{eq-Sigmaf-inf}) at ${\cal O}(f^{1})$.

However, we will get additional divergences from cross terms at ${\cal O}(c^{1}f^{1})$, which have the
same natural order. There are two diagrams with one $c$ and $f$ insertion each on the internal line
(corresponding to the two orders in which the insertions may appear). The first such diagram (with $c$ then
$f$ along the direction of fermion number flow) yields
\begin{eqnarray}
-i\Sigma_{cf} & = &  (-ig)^{2}\int\frac{d^{d}k}{(2\pi)^{d}}\frac{i(\slashed{k}+\slashed{p}+m)}{(k+p)^{2}-m^{2}}
\left[ic^{\nu\mu}\gamma_{\nu}(k+p)_{\mu}\right]\frac{i(\slashed{k}+\slashed{p}+m)}{(k+p)^{2}-m^{2}} \nonumber\\
& & \times\left[-\gamma_{5}f^{\rho}(k+p)_{\rho}\right]
\frac{i(\slashed{k}+\slashed{p}+m)}{(k+p)^{2}-m^{2}}\frac{i}{k^{2}-\mu^{2}} \\
& = & -ig^2\int\frac{d^{d}k}{(2\pi)^{d}}\frac{\gamma_{5}c^{\nu\mu}f^{\rho}(\slashed{k}+\slashed{p}-m)
\gamma_{\nu}(k+p)_{\mu}(k+p)_{\rho}}{[(k+p)^{2}-m^{2}]^{2}(k^{2}-\mu^{2})}
\end{eqnarray}
Proceeding as usual, we get 
\begin{equation}
-i\Sigma_{cf}=-ig^{2}\gamma_{5}c^{\nu\mu}f^{\rho}\int\! dx\,2x\!\int\!\frac{d^{d}l}{(2\pi)^{d}}
\frac{[\slashed{l}+(1-x)\slashed{p}-m]\gamma_{\nu}[l+(1-x)p]_{\mu}[l+(1-x)p]_{\rho}}
{(l^{2}-\Delta)^{3}}.
\end{equation}

The contribution from the $f$-then-$c$ diagram similarly turns out to be
\begin{equation}
-i\Sigma_{fc}=-ig^{2}\gamma_{5}c^{\nu\mu}f^{\rho}\int\! dx\,2x\!\int\!\frac{d^{d}l}{(2\pi)^{d}}
\frac{\gamma_{\nu}[\slashed{l}+(1-x)\slashed{p}+m][l+(1-x)p]_{\mu}[l+(1-x)p]_{\rho}}
{(l^{2}-\Delta)^{3}}.
\end{equation}
Taking the sum of the two contribution and simplifying yields
\begin{equation}
-i\Sigma_{cf+fc}=
-4ig^{2}\gamma_{5}c^{\nu\mu}f^{\rho}\int dx\,x\int\frac{d^{d}l}{(2\pi)^{d}}
\frac{[l+(1-x)p]_{\nu}[l+(1-x)p]_{\mu}[l+(1-x)p]_{\rho}}{(l^{2}-\Delta)^{3}}.
\end{equation}
The innermost integral is now identical with the one in (\ref{eq-Sigmafff-int}). Thus the simplification
(\ref{eq-Nfff}) of the numerator is applicable [although (\ref{eq-Nfff-symm}) is not,
since the expression
is not being contracted with the totally symmetric $f^{\mu}f^{\nu}f^{\rho}$, but rather with
$c^{\nu\mu}f^{\rho}$]. Therefore, the final ${\cal O}(c^{1}f^{1})$ fermion self-energy reduces to
\begin{equation}
-i\Sigma_{cf+fc}\overset{{\rm LV}}{\sim}\frac{g^{2}}{6}\left[\gamma_{5}
\left(c^{\nu}\,_{\nu}f^{\rho}p_{\rho}+2c^{\nu\mu}f_{\nu}p_{\mu}\right)\right]\eta.
\end{equation}
This has the structure of a potential radiative correction to $f$, but it has a more intricate form
than has previously been encountered.
The effective $f^{\mu}$ would receive
a contribution proportional to $c^{\nu\mu}f_{\nu}$, which does not necessarily point along
the same spacetime direction as $f$ itself.

However, we do note one very interesting feature of the self-energy at this order. The full
self-energy contribution
with both ${\cal O}(f^{3})$ and ${\cal O}(c^{1}f^{1})$ terms
(not just their divergent parts) may be cast in the form
\begin{eqnarray}
-i\Sigma_{fff+cf+fc} & = & -4ig^{2}\gamma_{5}\left(c^{\nu\mu}-\frac{1}{2}f^{\nu}f^{\mu}\right)f^{\rho}
\int dx\,x(1-x) \nonumber\\
& & \times\int\frac{d^{d}l}{(2\pi)^{d}}\frac{l_{\mu}l_{\nu}p_{\rho}+l_{\mu}l_{\rho}p_{\nu}
+l_{\nu}l_{\rho}p_{\mu}+(1-x)^{2}p_{\mu}p_{\nu}p_{\rho}}{(l^{2}-\Delta)^{3}}.
\end{eqnarray}
This has the elegant consequence that when the physically observable Lorentz violation coefficient
tensor $2c^{\nu\mu}-f^{\nu}f^{\mu}$ vanishes, then these higher-order radiative corrections vanish
as well.

\section{Conclusion}
\label{sec-concl}

The central point of this paper is that in the SME, which is an EFT for describing Lorentz and
CPT violation, there may be explicit renormalization scheme dependence in the RG $\beta$-functions.
Moreover,
this scheme dependence already occurs at one-loop order. Depending on what is more calculationally
convenient, it may be preferable to use a prescription that yields vanishing $\beta_{f}$, or one
in which $c^{\nu\mu}$ and $f^{\nu}f^{\mu}$ have the same RG evolution. However, as seen in
(\ref{eq-betacff}), which governs the scale dependence of the physically observable quantity
$2c^{\nu\mu}-f^{\nu}f^{\mu}$ that
appears in the fermion kinetic energy, physical predictions should not depend on the choice of
scheme; we are talking about differences in accounting, not physics.

The reason for the scheme dependence in the RG functions is ultimately that the underlying action
for the theory contains redundancies in its parameterization. The minimal SME Lagrange density includes
all the superficially renormalizable terms that it is possible to write down involving Dirac matrices and
derivatives acting on standard model fields. However, not all parameters in the Lagrange density are
physically distinguishable. In particular, the same physics can be described with either a $c$ term or a
$f$ term---or an intermediate combination of both. When quantum corrections are included, this ambiguity
naturally persists. A physical
effect that occurs due to virtual particle interactions may be similar to the effect of a tree-level $f$
term, so it may make sense to treat the quantum modification as a radiative correction to $f$ itself.
However, since the effects of a field operator with the Lorentz structure of a $f$ term---whether at tree
level or radiatively generated---cannot be distinguished from the effects of a $c$, any radiatively
induced contribution to $f$ could be alternatively interpreted as a radiative correction to $c$.

The explicit one-loop $\beta$-functions (\ref{eq-betacX}--\ref{eq-betafX}) show how the the radiative
corrections to a Lorentz-violating fermion kinetic term can be parceled out as contributions to either
$c$ or $f$. A more cumbersome alternative way of demonstrating the existence of the ambiguity also exists.
It is possible, beginning with a bare theory containing both $c$ and $f$, to use a transformation like
(\ref{eq-redefgeneral}) to rotate all the Lorentz violation into the bare $c$ term. Then the quantum
corrections can be calculated and any radiative corrections to $c$ determined, before performing
another rotation in the Dirac space, to convert the
renormalized $c$ operator into an appropriate combination of renormalized $c$ and $f$ terms.

As previously noted, we chose in this paper to work with a Yukawa theory (with no
explicit Lorentz-violating
terms appearing in the fermion-boson vertex) purely for reasons of simplicity. There does not seem to
be any reason to expect that the resolution we found for the puzzle concerning $\beta_{f}$ will not apply
more generally, including to the gauge sectors of the SME.
Obviously, the accounting of Feynman diagrams in a gauge theory---in which the $c$ and $f$ terms
appear in the vertex as well as the fermion propagator---will be quite a bit more intricate than in the
Yukawa theory considered here. However, we anticipate that the interplay between
$c$ and $f$ should remain qualitatively the same.

It would nonetheless be interesting to understand
the details of this interplay in more general Lorentz-violating theories. Extension of the these results
to the gauge sector is one obvious area where further research is possible, but there are
also questions still to be answered in the Yukawa sector of the SME. When other SME terms are
present in the Yukawa action (in either the fermion
propagation sector or in the interaction vertex), the $f$ may mix
nonlinearly with these additional coefficients. The general pattern of scheme dependence in the RG
structure should persist in these more general SME Yukawa theories, but the precise details of which terms
are involved remain to be worked out.
Answering these various questions can provide further insights into the structure of the SME, as well as
nonlinear regimes in QFT more generally.

\end{document}